\documentclass[%
 amsmath,amssymb,
 aps, physrev,
]{revtex4-2}

\usepackage{graphicx}
\usepackage{dcolumn}
\usepackage{bm}
\usepackage{braket}
\usepackage{subfig}
\usepackage[normalem]{ulem}
\usepackage{xcolor}
\definecolor{Mahogany}{RGB}{192, 64, 0}
\usepackage[colorlinks=true, linkcolor=Mahogany, citecolor=Mahogany, urlcolor=Mahogany]{hyperref}

\begin{document}


\title{\textbf{Analytical Expressions for Neutrino Oscillation} 
}%

\author{Adriano Cherchiglia}
\affiliation{Instituto de Física Gleb Wataghin, Universidade Estadual de Campinas \\ Rua Sérgio Buarque de Holanda, 777, Campinas, SP, Brasil}%
\author{Macello Jales}
\affiliation{Instituto de Física Gleb Wataghin, Universidade Estadual de Campinas \\ Rua Sérgio Buarque de Holanda, 777, Campinas, SP, Brasil}%
\author{Guilherme Nogueira}
\affiliation{Instituto de Física Gleb Wataghin, Universidade Estadual de Campinas \\ Rua Sérgio Buarque de Holanda, 777, Campinas, SP, Brasil}%
\author{Maressa P. Sampaio}
\affiliation{Instituto de Física Gleb Wataghin, Universidade Estadual de Campinas \\ Rua Sérgio Buarque de Holanda, 777, Campinas, SP, Brasil}%
\author{Pedro C. de Holanda}
\affiliation{Instituto de Física Gleb Wataghin, Universidade Estadual de Campinas \\ Rua Sérgio Buarque de Holanda, 777, Campinas, SP, Brasil}%


\date{\today}

\begin{abstract}
Research in neutrino physics has been very active, both in experimental advances, with a new generation of detectors in operation and planning, and in theoretical discussions regarding the fundamental nature of the neutrino. This scientific dynamism has attracted many new students to the field. One of the first topics studied in neutrino physics by newcomers is the formalism of neutrino flavor oscillations and its associated phenomenology. We present this work as a compilation of this basic knowledge, through a step-by-step approach that facilitates an efficient understanding of this vast theoretical and experimental landscape.

\end{abstract}

\maketitle

\tableofcontents

\newpage
\section{Introduction}
The Standard Model of Particle Physics (SM) was firmly established in the 70's, encompassing three of the four fundamental interactions observed in Nature. In this framework, apart from the gauge bosons related to the electromagnetic and strong interactions, there was only a particle considered massless: the neutrino. Contrary to the gauge bosons, whose massless nature is directly connected to the conserved symmetries of the SM, there was no \textit{a priori} theoretical reason for massless neutrino. The main reason was phenomenological. 

This picture remained untouched until the end of the 90's and begin of 2000's, when there was solid confirmation of the phenomenon of neutrino oscillation, both by Super-Kamiokande~\cite{Super-Kamiokande:1998kpq} and the Sudbury Neutrino Observatory (SNO) ~\cite{SNO:2001kpb}. This phenomenon can only be explained if neutrinos are massive particles, contradicting the hypothesis included in the SM. This fact can be regarded as one of the few solid experimental observations that require the SM to be extended.  

A particular interesting question is how to accommodate a massive neutrino within the Higgs mechanism.  This is only possible if neutrinos with right-handed chirality are introduced. However, since their electroweak hypercharge is null, they do not couple with the neither the photon or the Z boson. They also do not couple with the W boson, since charged currents are left-handed. Therefore, in the framework of the SM particles, they could only couple with the Higgs. Since neutrinos masses are tiny, this coupling would be also very small, extremely challenging to be detected by present and future experiments. Nevertheless, other mechanisms for generating neutrinos masses have been put forward, which require the inclusion of other particles beyond the SM (see, for instance, \cite{Cai:2017jrq} for a review).

Apart from the mechanism to generate their masses, neutrinos hide other mysteries. In particular, they are the only particles in the SM that could be their own antiparticles (Majorana), since they do not carry electromagnetic (or colour) charge. On-going experiments on neutrinoless double beta decay aim to unveil this characteristic~\cite{DellOro:2016tmg,Dolinski:2019nrj}. Another intriguing question is whether CP violation can also occur in the leptonic sector. Current and near-future oscillation neutrino experiments aim to provide a definite answer for this question. At present, the CP conservation seems to be disfavored, but more date is needed.  

Given this general picture, the study of neutrinos properties presents itself as a promising path towards an extended SM. Therefore, precise knowledge of the neutrino oscillation phenomenon is desirable, motivating next generation experiments such as DUNE, Hyper-Kamiokande and JUNO. In a nutshell, the differential event rate for neutrinos of flavor $\beta$ with energy $E_\nu$ to be detected at a distance $L$ from the source $S$, where they were produced with flavor $\alpha$, is given by 
\begin{equation}
\label{eq:rate}
  R_{\alpha\beta}^S = N_T\Phi_\alpha(E_{\nu})\sigma_\beta(E_{\nu})P_{\alpha\beta}
\end{equation}
where $N_{T}$ is the number of target particles, and $\Phi_\alpha(E_{\nu})$, $\sigma_\beta(E_{\nu})$ are the incident flux and detection cross-section, respectively. Particularly important are the neutrino conversion probabilities $P_{\alpha\beta}$, which are the main topic of our work. We will review their definition in sec.II, providing analytic expressions both in vacuum and in the modifications of the neutrino parameters in matter (sec.III). A series of approximations will be discussed, as well as their limits of applicability. We aim to provide the reader with easy-to-use expressions, helping into developing their physical intuition regarding the phenomenon of neutrino oscillation.

\section{Vacuum Oscillation Probabilities}

The usual expression for the neutrino flavor probabilities, evaluated within the framework of quantum mechanics, can be written as~\cite{Giunti:2007ry}
\begin{eqnarray}
P_{\alpha\beta}=\delta_{\alpha\beta}&-&4\sum_{k>j}\Re \left[U^*_{\alpha k}U_{\beta k}U_{\alpha j}U^*_{\beta j}\right]\sin^2\left(\frac{\Delta m^2_{kj}L}{4E}\right) \nonumber \\
&+&2\sum_{k>j}\Im\left[U^*_{\alpha k}U_{\beta k}U_{\alpha j}U^*_{\beta j}\right]\sin\left(\frac{\Delta m^2_{kj}L}{2E}\right). \label{Prob}
\end{eqnarray}
The dependence on neutrino masses in Eq.~(\ref{Prob}) is encompassed by the difference in the squares of the individual mass eigenstates, $\Delta m_{kj}^{2} = m_{k}^{2} - m_{j}^{2}$. Meanwhile, the ratio of the distance between the neutrino source and the detector, $L$, by the neutrino energy $E$ is known as the baseline. Additionally,  $U$ is the PMNS matrix, which in its common parametrization~\cite{ParticleDataGroup:2024cfk} takes the following form
\begin{eqnarray}
U&=& R(\theta_{23})R(\theta_{13}, \delta_{CP})R(\theta_{12})\label{PMNS}=\left(
\begin{array}{ccc}
c_{12}c_{13} & s_{12}c_{13} & s_{13}e^{-i\delta_{CP}} \\
-s_{12}c_{23}-c_{12}s_{23}s_{13}e^{i\delta_{CP}} & c_{12}c_{23}-s_{12}s_{23}s_{13}e^{i\delta_{CP}} & s_{23}c_{13} \\
s_{12}s_{23}-c_{12}c_{23}s_{13}e^{i\delta_{CP}} & -c_{12}s_{23}-s_{12}c_{23}s_{13}e^{i\delta_{CP}} & c_{23}c_{13} 
\end{array}
\right),
\label{eq:PMNS}
\end{eqnarray}
where, we are employing the notation $s_{ij} = \sin \theta_{ij}$ and $c_{ij} = \cos \theta_{ij}$. The component $R(\theta_{ij}, \delta_{ij})$ is a $3 \times 3$ matrix that represents a rotation in the $ij$-plane by an angle $\theta_{ij}$, with their respective CP phases. The non-trivial entries can be explicitly written as~\cite{deGouvea:2022kma}
\begin{eqnarray}
R(\theta_{ij}) = 
\begin{pmatrix}
c_{ij} & s_{ij}  \\
-s_{ij}  & c_{ij}
\end{pmatrix}; \hspace{0.6cm}
R(\theta_{ij}, \delta_{ij}) = 
\begin{pmatrix}
c_{ij} & s_{ij} e^{-i\delta_{ij}} \\
-s_{ij} e^{i\delta_{ij}} & c_{ij}
\end{pmatrix}.
\label{eq:rotation}
\end{eqnarray}
 This matrix is essential to comprehend how neutrino mixing works. One important feature of the $U$ matrix is that it is unitary ($U^{\dagger}U = I$), and this leads to the conservation of the transition probability, which will be employed later.

Given the hierarchical characteristic of the neutrino parameters, there are a number of approximations that are commonly used. Here we present in a concise way the main ones. 

\subsection{Large mass scale (atmospheric)}


After adding the brief comment about the vacuum oscillation probability, we are able to explore some limits that can be used in the context of the advancement of Eq.~(\ref{Prob}). Now,
for $L=0$ we obviously get $P_{\alpha\beta}=\delta_{\alpha\beta}$. As $L$ increases, the effects of the larger $\Delta m^2_{ij}$ will show up first, and since the mass differences are hierarchical, we may start from the regime where:
\[\frac{\Delta m^2_{31}L}{4E} =  \frac{\Delta m^2_{32}L}{4E}\gtrsim 1~~~;~~~ \frac{\Delta m^2_{21}L}{4E} = 0.
\]

Then, for a 3-neutrino scenario, the probability is
\begin{eqnarray} \label{general-prob}
P_{\alpha\beta}=\delta_{\alpha\beta}&-&4\left(\sum_{j<3}\Re \left[U^*_{\alpha 3}U_{\beta 3}U_{\alpha j}U^*_{\beta j}\right]\right)\sin^2\left(\frac{\Delta M^2L}{4E}\right) \nonumber\\
&+&2\left(\sum_{j<3}\Im\left[U^*_{\alpha 3}U_{\beta 3}U_{\alpha j}U^*_{\beta j}\right]\right)\sin\left(\frac{\Delta M^2L}{2E}\right),
\end{eqnarray}
where $\Delta M^2 = \Delta m^2_{31} = \Delta m^2_{32}$.

\vspace{0.5cm}
{\parindent=0pt \bf Reactor experiments:}
\vspace{0.2cm}

We now want to employ Eq.~(\ref{general-prob}) in experimental scenarios. For instance, in a nuclear reactor, the beta decay of fission products is an abundant source of\,
$\overline{\nu}_e$. Thus, reactor experiments \cite{suekane2016double}, e.g. Double-Chooz, Daya-Bay and Reno, usually study the survival probability of electron anti-neutrinos  
\begin{eqnarray*}
P_{ee}&=&1-4\left(|U_{e3}|^2|U_{e1}|^2 + |U_{e3}|^2|U_{e2}|^2 \right)\sin^2\left(\frac{\Delta M^2L}{4E}\right)\\
&=&1-4|U_{e3}|^2\left(1-|U_{e3}|^2\right)\sin^2\left(\frac{\Delta M^2L}{4E}\right),
\end{eqnarray*}
which, replacing the mixing angles given by Eq.~(\ref{eq:PMNS}), it is possible to achieve the familiar form
\begin{equation}
   P_{ee}=1-\sin^2(2\theta_{13})\sin^2\left(\frac{\Delta M^2L}{4E}\right).
   \label{Pee-reactor}
\end{equation}

There are no reactor experiments measuring a non-electronic appearance in an electronic neutrino flux, but the probabilities can be calculated by the same procedure. However, there is a more elegant way to arrive at these probabilities. First, consider the time evolution equation
\[
i\frac{d}{dt}\left(\begin{array}{c}\nu_e\\\nu_\mu\\\nu_\tau\end{array}\right) = 
H\left(\begin{array}{c}\nu_e\\\nu_\mu\\\nu_\tau\end{array}\right),
\]
which in mass basis, can be written as
\[
i\frac{d}{dt}\left(\begin{array}{c}\nu_1\\\nu_2\\\nu_3\end{array}\right) = 
\left(
\begin{array}{ccc}
\frac{m_1^2}{2E} & 0 & 0 \\
0 & \frac{m_2^2}{2E} & 0 \\
0 & 0 & \frac{m_3^2}{2E} \\
\end{array}
\right)
\left(\begin{array}{c}\nu_1\\\nu_2\\\nu_3\end{array}\right).
\]
To return to flavour basis, we observe that the mixing between mass eingenstates $\nu_i$ and flavour eigenstates $\nu_\alpha$ is given by $\nu_\alpha = U \nu_i$. Also, we can decompose the $ U $ matrix into its rotational components, and following the notation defined in Eq.~(\ref{PMNS}), we adopt the prescription $ U_{ij} \equiv R(\theta_{ij})$, resulting in the following compact parametrization:

\[
i\frac{d}{dt}\left(\begin{array}{c}\nu_e\\\nu_\mu\\\nu_\tau\end{array}\right) = 
U_{23}U_{13}U_{12}
\left(
\begin{array}{ccc}
\frac{m_1^2}{2E} & 0 & 0 \\
0 & \frac{m_2^2}{2E} & 0 \\
0 & 0 & \frac{m_3^2}{2E} \\
\end{array}
\right)
U_{12}^\dagger U_{13}^\dagger U_{23}^\dagger 
\left(\begin{array}{c}\nu_e\\\nu_\mu\\\nu_\tau\end{array}\right).
\]
When $\Delta m^2_{21}=m_2^2-m_1^2=0$, the $U_{12}$ and $U_{12}^\dagger$ are absorbed as internal rotations that cancel out. Then, one can define the following basis
\begin{equation}
\left(\begin{array}{c} \nu_\mu'\\\nu_\tau'\end{array}\right) =
U_{23}^\dagger \left(\begin{array}{c} \nu_\mu\\\nu_\tau\end{array}\right),
\label{eq:nuprime}
\end{equation}
and rewrite the evolution equation as
\[
i\frac{d}{dt}\left(\begin{array}{c}\nu_e\\\nu_\mu'\\\nu_\tau'\end{array}\right) = 
U_{13}
\left(
\begin{array}{ccc}
\frac{m^2}{2E} & 0 & 0 \\
0 & \frac{m^2}{2E} & 0 \\
0 & 0 & \frac{M^2}{2E} \\
\end{array}
\right)
U_{13}^\dagger
\left(\begin{array}{c}\nu_e\\\nu_\mu'\\\nu_\tau'\end{array}\right),
\]
where $m=m_1=m_2$ and $M=m_3$. The second family decouples since $U_{13}$ and $U_{13}^\dagger$ do not affect it, so the evolution equation reduces to
\[
i\frac{d}{dt}\left(\begin{array}{c}\nu_e\\\nu_\tau'\end{array}\right) = 
U_{13}
\left(
\begin{array}{ccc}
\frac{m^2}{2E} & 0 \\
0 & \frac{M^2}{2E} \\
\end{array}
\right)
U_{13}^\dagger
\left(\begin{array}{c}\nu_e\\\nu_\tau'\end{array}\right).
\]

The survival probability $P_{ee}$ was already evaluated in Eq~(\ref{Pee-reactor}), and we know that the sum of these two probabilities (survival and oscillation to $\tau'$) must add to one, so
\begin{equation}\label{p_e-tau}
    P_{e\tau'}=\sin^2(2\theta_{13})\sin^2\left(\frac{\Delta M^2L}{4E}\right),
\end{equation}

and since by Eq.~\ref{eq:nuprime} we have $\braket{\nu_\mu|\nu_\tau'}=s_{23}$ and $\braket{\nu_\tau|\nu_\tau'}=c_{23}$, we can write: 
\[
P_{e\mu}=s^2_{23}\sin^2(2\theta_{13})\sin^2\left(\frac{\Delta M^2L}{4E}\right)
~~;~~~
P_{e\tau}=c^2_{23}\sin^2(2\theta_{13})\sin^2\left(\frac{\Delta M^2L}{4E}\right).
\]
To sum up, the oscillation of electron anti-neutrinos driven by the large mass scale occurs to an admixture of muon and tau neutrinos, with weights given by $\theta_{23}$.

\vspace{0.5cm}
{\parindent=0pt \bf Accelerator experiments:}
\vspace{0.2cm}

Various accelerator experiments \cite{nath2021detection}, such as Minos~\cite{MINOS:2020llm} and T2K~\cite{T2K:2023mcm}, produce beams of $\nu_\mu$ or $\overline{\nu}_\mu$ by accelerating and colliding protons into a target. Likewise, cosmic rays (usually protons) interact with other nuclei in the atmosphere, mostly producing muonic (anti-)neutrinos \cite{kajita2010atmospheric}, which are detected by experiments such as IceCube~\cite{IceCubeCollaboration:2023wtb} and Super-Kamiokande~\cite{Super-Kamiokande:2017yvm,T2K:2024wfn}. In these contexts, the survival probability $P_{\mu \mu}$ is useful and can be derived from Eq.~(\ref{general-prob}):
\begin{eqnarray}\label{p_mu-mu1}
P_{\mu\mu}&=&1-4|U_{\mu 3}|^2\left(1-|U_{\mu 3}|^2\right)\sin^2\left(\frac{\Delta M^2L}{4E}\right) \nonumber \\
&=&1-4c^2_{13}s^2_{23}(1-c^2_{13}s^2_{23})\sin^2\left(\frac{\Delta M^2L}{4E}\right) \nonumber \\
\label{p_mu-mu}
&=&1-(s^2_{23}\sin^2(2\theta_{13})+\sin^2(2\theta_{23})c^4_{13})\sin^2\left(\frac{\Delta M^2L}{4E}\right).
\end{eqnarray}
Using the unitarity property of the PMNS matrix, the rows (or columns) of the matrix must satisfy the orthogonality conditions
\[
\sum_k U_{\mu k} U^*_{e k} = 0,
\]
which can be replaced in Eq.~(\ref{p_mu-mu1}). This makes it possible to rewrite the appearance probabilities as
\begin{eqnarray*}
P_{\mu e}&=&4|U_{\mu 3}|^2|U_{e3}|^2\sin^2\left(\frac{\Delta M^2L}{4E}\right)=s^2_{23}\sin^2(2\theta_{13})\sin^2\left(\frac{\Delta M^2L}{4E}\right),\\
P_{\mu \tau}&=&4|U_{\mu 3}|^2|U_{\tau 3}|^2\sin^2\left(\frac{\Delta M^2L}{4E}\right)=c^4_{13}\sin^2(2\theta_{23})\sin^2\left(\frac{\Delta M^2L}{4E}\right).
\end{eqnarray*}
which are related to the second and third term in Eq.~(\ref{p_mu-mu}). So the main oscillation channel of atmospheric and accelerator muon neutrinos is to tau neutrinos~\cite{Super-Kamiokande:2017edb}, with a small production of electron anti-neutrinos driven by the size of $\theta_{13}$.

\subsection{Small mass scale (solar)}


The next approximation to be taken into account is when the solar mass scale does not vanish, but have a small value. Indeed, this limit can be defined through the relation
\[
\frac{\Delta m^2_{21}L}{4E}\gtrsim 1.
\]
In this regime, we can write the probability of survival of electron neutrinos as a simple expansion of Eq.~(\ref{Prob}), where the term with the solar scale is fully expressed as
\[
P_{ee}=1-4|U_{e2}|^2|U_{e1}|^2\sin^2\left(\frac{\Delta m^2_{21}L}{4E}\right)+f,
\]
while the 
large mass scale contributions are grouped together in the following parameter:
\[
f=-4|U_{e3}|^2\left[|U_{e1}|^2\sin^2\left(\frac{\Delta m^2_{31}L}{4E}\right)+|U_{e2}|^2\sin^2\left(\frac{\Delta m^2_{32}L}{4E}\right)\right].
\]
In this small mass scale regime, we can average out the oscillation driven by the larger terms $\Delta m^2_{31}L/{4E}$ and $\Delta m^2_{32}L/4E$, resulting in
\[
f\sim -2|U_{e3}|^2\left(|U_{e1}|^2+|U_{e2}|^2\right)
=-2|U_{e3}|^2\left(1-|U_{e3}|^2\right),
\]
which, by replacing the explicit values of the $U$ matrix elements with their corresponding mixing angles, makes it possible to derive, after some algebraic manipulation, the electron neutrino survival probability:
\[
P_{ee}=c^4_{13}P_{ee}^{(2fam)}+s^4_{13},
\]
where $P_{ee}^{(2fam)}$ is the survival probability in the limit $\theta_{13}\rightarrow 0$:
\[
P_{ee}^{(2fam)}= 1-\sin^2(2\theta_{12})\sin^2\left(\frac{\Delta m^2_{21}L}{4E}\right).
\]
The only approximation used on this formula regards the mass hierarchy. If we also want to disregard $\theta_{13}$, then the survival probability trivially reduces to the 2-family one.
This is the expression used for understanding KamLAND results~\cite{KamLAND:2010fvi}.


\section{Matter effects}

Although very useful to describe the oscillation of terrestrial neutrinos, astrophysical neutrinos are usually created in dense environments, such as the cores of stars or supernovae. In these conditions, matter effects significantly impact their oscillation behavior, as interactions with particles in the medium modify the effective mixing angles and mass differences. But even for terrestrial neutrinos, considering matter effects on flavor conversion probabilities are crucial when precision predictions are needed, such as in the case of determining the mass ordering of the neutrino families. 

The main effect of matter effects can be analyzed through a modification on the mass and mixing matrix parameters. If neutrinos travels through a constant matter environment, the formulas derived in previous sections could be replaced by the same expressions, but with the new values for mass and mixing angles calculated in matter. Finding expressions for these parameters is what we present in the following section. However, for a varying matter density, new phenomena, like resonant flavor conversion such as the MSW effect~\cite{Wolfenstein:1977ue, Mikheyev:1985zog} have to be considered, resulting in very different conversion probabilities compared to those in a vacuum. We will not focus on conversion probabilities in varying density environments in the present work.

The vacuum Hamiltonian ($H_{\text{vac}}$) is modified to the matter Hamiltonian ($H_{mat}$), which effectively accounts for the following displacement
\begin{equation}
  H_{mat} = H + V_{CC},
\end{equation}
where the matrix $V_{CC}$ encompass the matter potential effect. 

At first we present some generic expressions for two families, which can be used as an approximation in some specific scenarios for three families, presented in afterward. In what follows, unless explicitly stated otherwise, we 
 use for the oscillation parameters the values:
\begin{gather*}
\Delta m^2_{21} = 8\,10^{-5}~\textrm{eV}^2, \quad 
\Delta m^2_{31} = 2.5\,10^{-3}~\textrm{eV}^2; \\
\theta_{12} = 0.59, \quad 
\theta_{13} = 0.148, \quad 
\theta_{23} = 0.738; \\
E_\nu = 10\,\textrm{MeV}.
\end{gather*}

\subsection{Two families}

Analyzing the neutrino oscillation problem within the framework of only two families yields significant benefits, as it envelopes the fundamental aspects of the mechanism while maintaining a simplified structure. Given the  mixing matrix between mass eingenstates $\nu_i$ and flavour eigenstates $\nu_\alpha$ already disscused and the rotation matrix defined in Eq~(\ref{eq:rotation}), 
%
the Hamiltonian in flavour basis assuming that the first state in flavour base is electronic can be written as:
\begin{eqnarray}
 H_{mat}&=&
\frac{m_1^2+m_2^2}{4E}+
U 
\left(\begin{array}{cc}
-\frac{\Delta m^2}{4E_\nu} & 0 \\
0 & +\frac{\Delta m^2_{21}}{4E_\nu} \\
\end{array} \right)
U^{\dagger}
+
\left(\begin{array}{cc}
V_{CC} & 0  \\
0 & 0 
\end{array} \right) \nonumber \\
&=&\frac{m_1^2+m_2^2}{4E}+\frac{V_{CC}}{2}+
\left(\begin{array}{cc}
-\frac{\Delta m^2}{4E_\nu}c_{2\theta} + \frac{V_{CC}}{2} & \frac{\Delta m^2}{4E_\nu}s_{2\theta} \label{matterHami1} \\
\frac{\Delta m^2}{4E_\nu}s_{2\theta} & +\frac{\Delta m^2}{4E_\nu}c_{2\theta} - \frac{V_{CC}}{2} \\
\end{array} \right),
\label{matterHami}
\end{eqnarray}
where 
$\Delta m^{2} = \Delta m_{21}^{2}$. In addition, the notation $c_{2 \theta} = \cos 2\theta$ and $s_{2 \theta} = \sin 2\theta$ was employed. The eigenvalues of the non-diagonal matrix can be calculated through the usual diagonalization process. Furthermore, the terms proportional to identity in Eq.~(\ref{matterHami}) should be added to the full eigenvalues:
%

\begin{equation}
 \lambda_{1,2}=\frac{m_1^2+m_2^2}{4E}+\frac{V_{CC}}{2}\pm\sqrt{\left(\frac{\Delta m^2}{4E_\nu}c_{2\theta} - \frac{V_{CC}}{2}\right)^2+\left(\frac{\Delta m^2}{4E_\nu}s_{2\theta}\right)^2}.   
 \label{2feigen}
\end{equation}
We have to diagonalize this matrix, i.e. find the modifications in the mixing angles in matter that leads to:
\begin{equation}
  \tilde{U^\dagger}H\tilde{U}=\textrm{diag}(\lambda_1,\lambda_2),
  \label{eq:matterH}
\end{equation}
where $\tilde{U}$ has the same structure as in Eq.~(\ref{eq:PMNS}), but with angles modified by matter interactions.  The mixing angle in matter $\tilde\theta$ that reproduces $H_{m}$ is
\begin{equation}
 \tan(2\tilde\theta)=\frac{\frac{\Delta m^2}{4E_\nu} s_{2\theta}}{\frac{\Delta m^2}{4E_\nu}c_{2\theta} - \frac{V_{CC}}{2}}.   
 \label{tangent}
\end{equation}
It is possible to 
expand Eq.~\ref{2feigen} in $\rho$ for particular limits, in order to achieve analytical approximations for the mixing angle and eigenvalues.

\vspace{0.5cm}
{\parindent=0pt {\bf Low densities:} $V_{CC}\ll \frac{\Delta m^2}{4E_\nu}c_{2\theta},\frac{\Delta m^2}{4E_\nu}s_{2\theta}$}
\vspace{0.3cm}

In this regime, Eq.~(\ref{2feigen}) can be rewritten as
$$
\lambda_{1,2}\sim \frac{m_1^2+m_2^2}{4E}+\frac{V_{CC}}{2}\pm\left[\frac{\Delta m^2}{4E_\nu}-\frac{1}{2}V_{CC}c_{2\theta}\right],
$$
which leads to the following analytical expressions for the eigenvalues
$$
\lambda_1=\frac{m_1^2}{2E_\nu}+c_{\theta}^2V_{CC}~~~;~~~\lambda_2=\frac{m_2^2}{2E_\nu}+s_{\theta}^2V_{CC}.
$$

{\parindent=0pt {\bf High densities:} $V_{CC}\gg \frac{\Delta m^2}{4E_\nu}s_{2\theta},\frac{\Delta m^2}{4E_\nu}c_{2\theta}$}
\vspace{0.3cm}

\begin{figure}[t]
	\centering
    \includegraphics[width=1\textwidth]{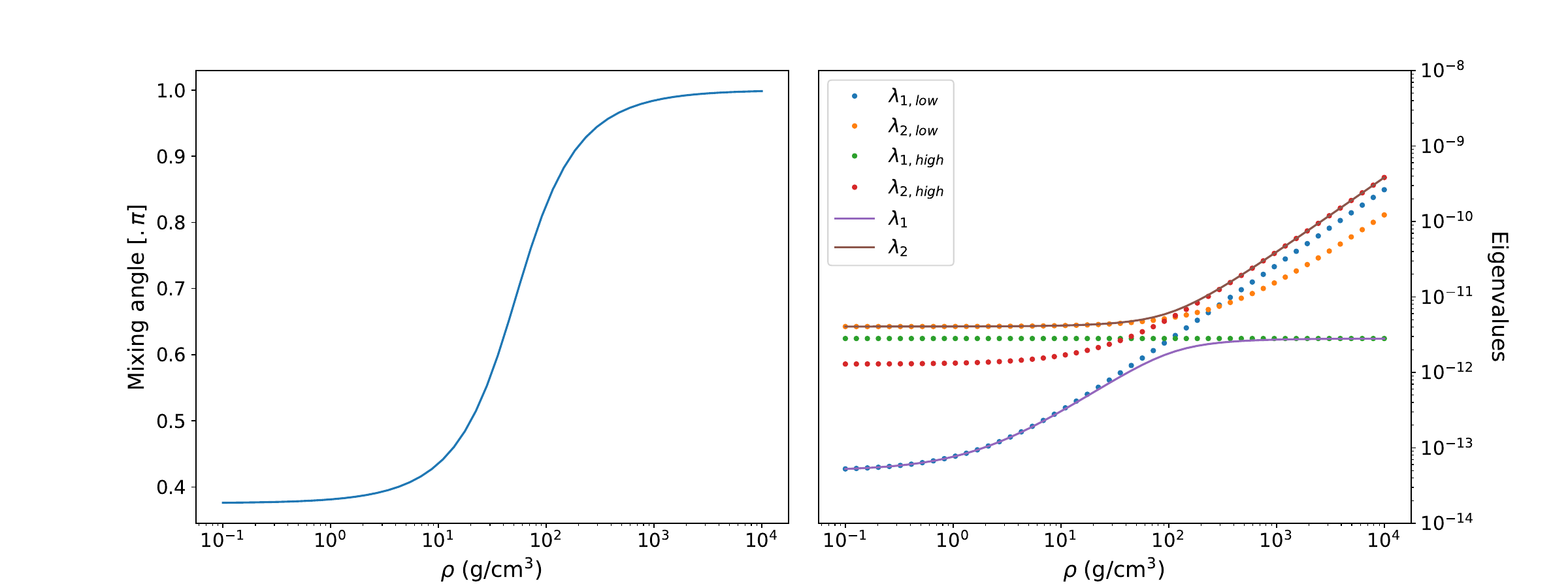}
    \caption{\raggedright Exact expressions (solid lines) and approximations (dotted) considering low and high densities for mixing angle (left panel) and eigenvalues (right panel).}
    \label{fig:2fam}
\end{figure}

The same prescription can be applied for the high matter density limit, where the eigenvalues are expanded in terms of $\rho$ leading to:
$$
\lambda_{1,2}\sim \frac{m_1^2+m_2^2}{4E}+\frac{V_{CC}}{2}\pm \left[\frac{V_{CC}}{2}-\frac{\Delta m^2}{4E_\nu}c_{2\theta}\right],
$$
which, once again, according to the ordering defined for the eigenvalues before, leads to analytical approximations:
$$
\lambda_1=\frac{m_2^2}{2E_\nu}-\frac{\Delta m^2}{2E_\nu}c_{\theta}^2.~~~;~~~\lambda_2=\frac{m_1^2}{2E_\nu}+\frac{\Delta m^2}{2E_\nu}s_{\theta}^2+V_{CC}.
$$
For a constant neutrino energy and varying density, we present in Fig.~(\ref{fig:2fam}) the eigenstates and mixing angle, together with the asymptotic behavior for low and high densities.

The behavior that is worth pointing out is that for high densities the first family eigenvalue equals the matter potential, and the mixing angle tends to $\pi$, which is a characteristic that can be found when expanding the analysis to 3 neutrino families in the next section.


\subsection{Three neutrino familes}

In spite of the two neutrino families analysis having a well-defined and simplified description, we will work on a more realistic case, which considers the neutrinos three families. To simplify our analisys we will set $\delta_{CP}=0$ in what follows. For this specific case, the explicit form of the Hamiltonian in the flavor basis can be written as
\begin{equation}
 H=U_{23}U_{13}U_{12} 
\left(\begin{array}{ccc}
\frac{m^2_1}{2E_\nu} & 0 & 0 \\
0 & \frac{m^2_2}{2E_\nu} & 0 \\
0 & 0 & \frac{m^2_3}{2E_\nu}
\end{array} \right)
U_{12}^{\dagger}U_{13}^{\dagger}U_{23}^{\dagger}
+
\left(\begin{array}{ccc}
V_{CC} & 0 & 0 \\
0 & 0 & 0\\
0 & 0 & 0
 \end{array} \right).
\label{eq:hmat}
\end{equation}


The Hamiltonian in Eq.~(\ref{eq:hmat}) can be analyzed using the same approach as in Eq.~(\ref{matterHami}), previously employed for the two-family regime. However, the key distinction in the three-family case lies in the presence of an additional eigenvalue, arising from the increased dimensionality of the system. The analysis will explore the behavior of the system in the low, intermediate, and high-density limits.
\vspace{0.5cm}

{\parindent=0pt {\bf Low densities:} $V_{CC}\sin (2\theta_{13})< \frac{\Delta m^2_{31}}{4E_\nu}$, $\frac{\Delta m^2_{21}}{4E_\nu}$}
\vspace{0.5cm}

Since $V_{CC}$ only affects $H_{11}$, the mixing angle $\theta_{23}$ can be rotated out redefining $\nu_\alpha'=U_{23}^{\dagger}\nu_\alpha$.
If we also rotate out $\theta_{13}$, defining a $\nu_\alpha''=U_{13}^{\dagger}\nu_\alpha'=U_{13}^{\dagger}U_{23}^{\dagger}\nu_\alpha$ we get:

\begin{eqnarray}
H^{''} = \frac{m_1^2+m_2^2}{4E_{\nu}} +
\left(\begin{array}{ccc}
-\delta_{21}c_{2\theta_{12}} & \delta_{21}s_{2\theta_{12}} & 0 \\
\delta_{21}s_{2\theta_{12}} & \delta_{21}c_{2\theta_{12}} & 0 \\
0 & 0 & \delta_{31}+\delta_{32}
\end{array} \right)
+ V_{CC}\left(\begin{array}{ccc}
c^2_{13} & 0 & s_{13}c_{13} \\
0 & 0 & 0 \\
s_{13}c_{13} & 0 & s^2_{13}
\end{array} \right).
 \label{H2prime}
\end{eqnarray}
where $\delta_{ji} = \Delta m^2_{ji} / 4E_\nu$.
For low densities we disregard the non-diagonal entry in the third term in right side of Eq.~\ref{H2prime}, decoupling our system into  2+1. 
It straightforward to obtain the third eigenvalue, related to the decoupled family:
\begin{eqnarray}
\lambda_3&=&\frac{m_3^2}{2E_\nu}+V_{CC}s^2_{13}.\label{eq:lambda3_low}
\end{eqnarray}
The remaining problem is solved by the diagonalization process of the non-diagonal subspace, which results in
\begin{eqnarray}
\lambda_{1,2}&=&
\frac{m_1^2+m_2^2}{4E}+\frac{V_{CC}c^2_{13}}{2}\pm\sqrt{\left(\delta_{21}c_{2\theta_{12}} - \frac{V_{CC}c^2_{13}}{2}\right)^2+\left(\delta_{21}s_{2\theta_{12}}\right)^2}. \label{eq:lambda12_low}
\end{eqnarray} 
The reduced 2-dimensional subsystem has an analytical solution for the mixing angle, as already shown in Eq.~(\ref{eq:matterH}), which gives
\begin{equation}
\tan(2\tilde\theta_{12})=\frac{\frac{\Delta m^2}{4E_\nu} s_{2\theta_{12}}}{\frac{\Delta m^2}{4E_\nu}c_{2\theta_{12}} - \frac{V_{CC}}{2}c_{13}^2}.  \label{eq:theta12}
\end{equation}
This modification leads to the following change in the mixing matrix
\[
\nu_\alpha''=\tilde{U}_{12}\nu_i
~~~\rightarrow~~~
U=U_{23}U_{13}\tilde{U}_{12}.
\]

\begin{figure}[t]
	\centering
    \includegraphics[width=1\textwidth]{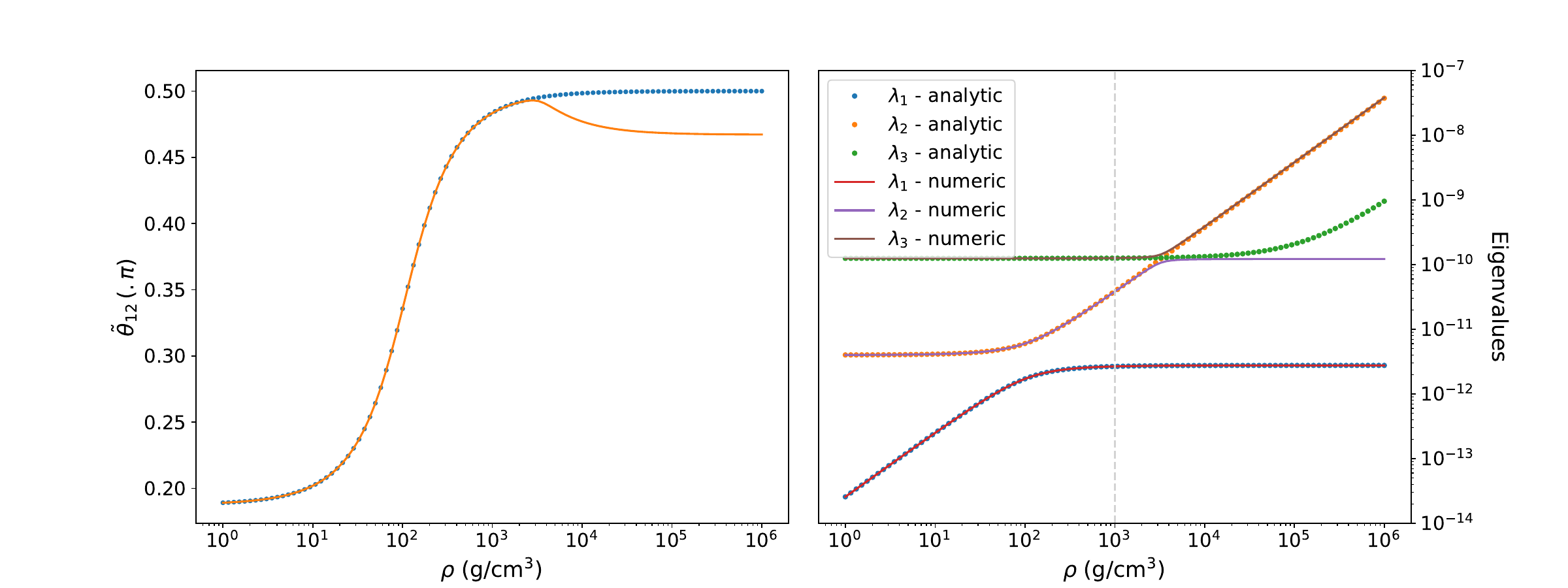}
     \caption{\raggedright Mixing angle $\theta_{12}$  (left panel) and eigenvalues $\lambda$'s (right panel), considering approximation for low densities (dotted) and numerical result (continuum line). The approximations are reliable up to $\rho\sim 10^3$ g/cm$^3$, marked as a vertical line.}
    \label{fig:lowrho}
\end{figure}

In Fig.~\ref{fig:lowrho} we present $\tilde{\theta}_{12}$ and eigenvalues as a function of density. The analytical expressions are reliable up to $\rho\sim10^3$ g/cm$^3$.
One interesting feature is that $\lambda_1$ provides a good approximation in the full range. The asymptotic value for large densities is
\begin{eqnarray*}
\lambda_1
&=&\frac{m_1^2+m_2^2}{4E}+\frac{V_{CC}c^2_{13}}{2}-\frac{V_{CC}c^2_{13}}{2}\sqrt{1-4\delta_{21}c_{2\theta_{12}}\frac{1}{V_{CC}c^2_{13}}+\left(\frac{2}{V_{CC}c^2_{13}}\delta_{21}\right)^2}\\
&\sim&
\frac{m_1^2}{2E}s^2_{12}+\frac{m_2^2}{2E}c^2_{12}.
\end{eqnarray*}


Since $\theta_{13}$ is small, this approximation should still be valid in the first resonance, $\tilde{\theta}_{12}=\pi/4$.

\vspace{0.5cm}
{\parindent=0pt {\bf intermediate to high densities:} $V_{CC}\sin (2\theta_{13})\gtrsim \frac{\Delta m^2_{31}}{4E_\nu} \gg \frac{\Delta m^2_{21}}{4E_\nu}$}

If we take $\Delta m^2_{21}=0$, we have:
\[
 H=
U_{23}U_{13}U_{12} 
\left(\begin{array}{ccc}
\frac{m^2_1}{2E_\nu} & 0 & 0 \\
0 & \frac{m^2_1}{2E_\nu} & 0 \\
0 & 0 & \frac{m^2_3}{2E_\nu}
\end{array} \right)
U_{12}^{\dagger}U_{13}^{\dagger}U_{23}^{\dagger}
+
\left(\begin{array}{ccc}
V_{CC} & 0 & 0 \\
0 & 0 & 0\\
0 & 0 & 0
 \end{array} \right).
\]
Now, besides absorbing $U_{23}$ in $\nu'$, we can perform the multiplication with $U_{12}$, arriving at:
\[
 H'=
U_{13}
\left(\begin{array}{ccc}
\frac{m^2_1}{2E_\nu} & 0 & 0 \\
0 & \frac{m^2_1}{2E_\nu} & 0 \\
0 & 0 & \frac{m^2_3}{2E_\nu}
\end{array} \right)
U_{13}^{\dagger}
+
\left(\begin{array}{ccc}
V_{CC} & 0 & 0 \\
0 & 0 & 0\\
0 & 0 & 0
 \end{array} \right) .
\]
Now is the second family that gets decoupled, and we can use the expressions for the 2-families. In this case, the eigenvalues and mixing angles can be respectively written as
\begin{equation}
\lambda_{2,3}=\frac{m_1^2+m_3^2}{4E}+\frac{V_{CC}}{2}\pm\sqrt{\left(\frac{\Delta m_{31}^2}{4E_\nu}c_{2\theta_{13}} - \frac{V_{CC}}{2}\right)^2+\left(\frac{\Delta m_{31}^2}{4E_\nu}s_{2\theta_{13}}\right)^2},
\label{eq:lambda23_high}
\end{equation}
\begin{equation}
\tan(2\tilde\theta_{13})=\frac{\frac{\Delta m_{31}^2}{4E_\nu} s_{2\theta_{13}}}{\frac{\Delta m_{31}^2}{4E_\nu}c_{2\theta_{13}} - \frac{V_{CC}}{2}} .   
\label{eq:theta13}
\end{equation}

It is interesting to note that the asymptotic value of $\lambda_3$ for low energies agrees with Eq.~(\ref{eq:lambda3_low}) at first order in $\rho$.

\begin{figure}[t]
	\centering
\includegraphics[width=1\textwidth]{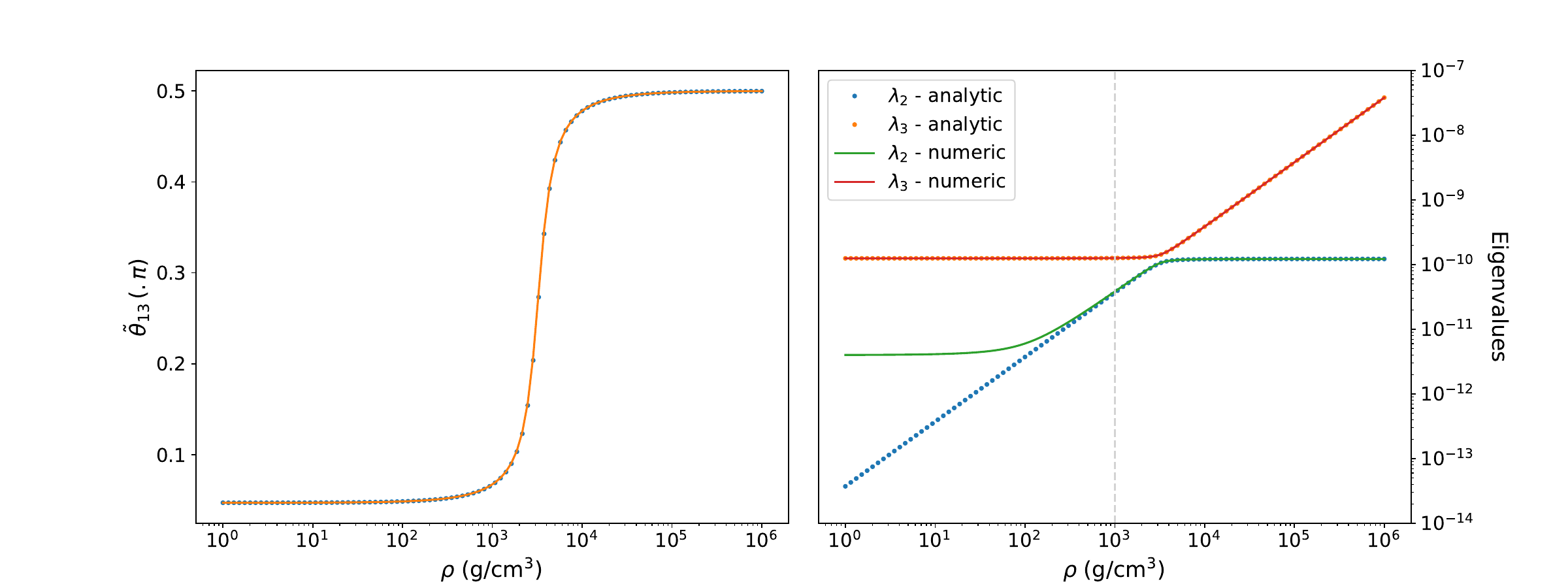}
\caption{ \raggedright
      Mixing angle $\theta_{13}$ (left panel) and eigenvalues $\lambda$'s (right panel), considering approximation for high densities (dotted) and numerical result (continuum line). The approximations are reliable starting at $\rho\sim 10^3$ g/cm$^3$, marked as a vertical line.}
\end{figure}

In summary, we have so far the following approximations:

\begin{itemize}
    \item $\lambda_1$ (Eq.~(\ref{eq:lambda12_low})) and $\lambda_3$ (Eq.~(\ref{eq:lambda23_high})) valid for all ranges of $\rho$.
    \item $\theta_{13}$ (Eq.~(\ref{eq:theta13})) also valid for all ranges of $\rho$.
    \item $\lambda_2$ for low densities (around first resonance) given by Eq.~(\ref{eq:lambda12_low}) and a different expression for high densities (around second resonance), Eq.~(\ref{eq:lambda23_high}).
    \item $\theta_{12}$ for low densities, given by Eq.~(\ref{eq:theta12}).

\end{itemize}

What is lacking is a expression for $\theta_{12}$ and $\theta_{23}$ for densities larger than the second resonance. Since we have reliable expressions for the other angles, it can be calculated as follows.

For very large densities $\tilde{\theta}_{13}\sim\pi/2$, and we can write
\[
\cos\tilde{\theta}_{13}\sim\epsilon~~~;~~~ \sin\tilde{\theta}_{13}\sim 1-\frac{\epsilon^2}{2},
\]
and the mixing matrix can be written in powers of $\epsilon$ as
\begin{eqnarray}
\tilde{U}=\left(
\begin{array}{ccc}
0 & 0 &  1\\
\multicolumn{2}{c}{\mathcal{O}} &
\begin{array}{c}
0\\
0
\end{array}
\end{array}
\right)
+ \epsilon
\left(
\begin{array}{ccc}
c_{12} & s_{12} &  0\\
\multicolumn{2}{c}{0} &
\begin{array}{c}
s_{23} \\
c_{23}
\end{array}
\end{array}
\right)
+\frac{\epsilon^2}{2}
\left(
\begin{array}{ccc}
0 & 0 & -1 \\
c_{12}s_{23} & s_{12}s_{23} & 0 \\
c_{12}c_{23} & s_{12}c_{23} & 0
\end{array}
\right),
\label{eq:Utilde}
\end{eqnarray}
where $\mathcal{O}$ can be written as a mixing matrix in 2 dimensions through a new angle $\alpha$:
\[
\mathcal{O}=
\left(\begin{array}{cc}
-s_{12}c_{23}-c_{12}s_{23} & c_{12}c_{23}-s_{12}s_{23} \\
s_{12}s_{23}-c_{12}c_{23} & -c_{12}s_{23}-s_{12}c_{23} 
\end{array}\right)
=
\left(\begin{array}{cc}
\cos\alpha & \sin\alpha \\
-\sin\alpha & \cos\alpha
\end{array}\right).
\]
Using Eq.(\ref{eq:hmat}) and the new parametrization in Eq.(\ref{eq:Utilde}), the  following relation
\[
\tilde{U}\textrm{diag}(\tilde{\lambda}_1,\tilde{\lambda}_2,\tilde{\lambda}_3)\tilde{U}^\dagger = U\textrm{diag}(\lambda_1,\lambda_2,\lambda_3)U^\dagger +V_{CC}.
\]
can be examined.
To first order in $\epsilon$ and considering $\tilde{\lambda}_{3}\gg \tilde{\lambda}_{1,2}$, the left side becomes:
\[
\frac{\tilde{\lambda}_1+\tilde{\lambda}_2}{2}+
\left(\begin{array}{ccc}
\frac{\Delta_{31}+\Delta_{32}}{2} & \epsilon\tilde{\lambda}_3s_{23}& \epsilon\tilde{\lambda}_3c_{23} \\
\begin{array}{c}
\epsilon\tilde{\lambda}_3s_{23} \\ \epsilon\tilde{\lambda}_3c_{23}
\end{array}
& \multicolumn{2}{c}{h_{2\times2}} 
\end{array}\right),
\]
where
\[
h_{2\times2}
=O
\left(\begin{array}{cc}
-\Delta_{21}/2 & 0  \\
0 & +\Delta_{21}/2
\end{array}\right)
O^\dagger
=
\frac{\Delta_{21}}{2}
\left(\begin{array}{cc}
-\cos2\alpha & \sin2\alpha \\
\sin2\alpha & \cos2\alpha
\end{array}\right),
\]
and $\Delta_{ji}=(\tilde{\lambda}_j-\tilde{\lambda}_i)/2$. 
Comparing with the left side of Eq.~(\ref{eq:hmat}), we can finally write:
\[
\tan\tilde{\theta}_{23}= (H_{mat})_{12}/(H_{mat})_{13}.
\]
Explicitly, after some algebraic manipulation, we obtain the following expression
\[
\tan\tilde{\theta}_{23}=\frac{(\Delta+\delta\cos2\theta_{12})\sin2\theta_{13}s_{23}+2\delta c_{13}\sin2\theta_{12}c_{23}}{(\Delta+\delta\cos2\theta_{12})\sin2\theta_{13}c_{23}-2\delta c_{13}\sin2\theta_{12}s_{23}}.
\]
\begin{figure}[t!]
    \centering
    \includegraphics[width=0.5\textwidth]{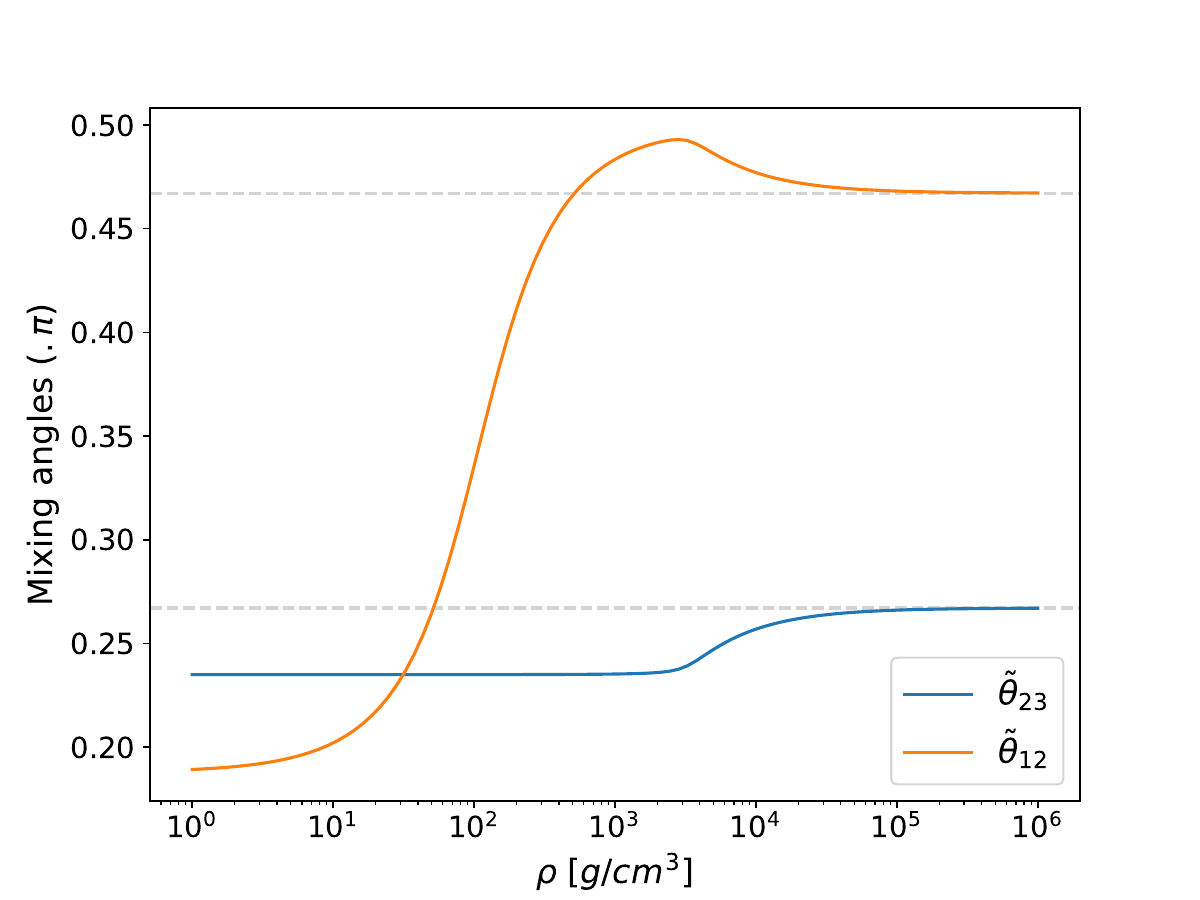}
    \caption{\raggedright Approximations considering high densities for $\theta_{23}$ and $\theta_{12}$. The asymptotic analytical expressions are marked as horizontal lines.}
    \label{fig:highrhot23}
\end{figure}

The relation between $\alpha$ and mixing angles $\theta_{12}$ and $\theta_{23}$ can be written realizing that:
\[
\mathcal{O}=
\left(\begin{array}{cc}
-s_{12}c_{23}-c_{12}s_{23} & c_{12}c_{23}-s_{12}s_{23} \\
s_{12}s_{23}-c_{12}c_{23} & -c_{12}s_{23}-s_{12}c_{23} 
\end{array}\right)
=
\left(\begin{array}{cc}
-s_{12} & c_{12} \\
-c_{12} & -s_{12}
\end{array}\right)
\left(\begin{array}{cc}
c_{23} & s_{23} \\
-s_{23} & c_{23}
\end{array}\right),
\]
so $\alpha$ can be related to a rotation of $\theta_{23}$ followed by a rotation of $\theta_{12}+\pi/2$. Then
\[
\tilde{\theta}_{12}=\alpha-\frac{\pi}{2}-\tilde{\theta}_{23},
\]
so, after finding $\tilde{\theta}_{23}$ it is straightforward to find the value of $\tilde{\theta}_{12}$. 
In Fig.~(\ref{fig:highrhot23}) we can see that the asymptotic value agrees with the numerical calculation.

%
%
%


\section{Conclusions}
In this article, we present analytical expressions for neutrino flavor conversion probabilities and mixing angles in matter across different experimental contexts. The matter dependence of the mixing angle $\theta_{23}$ is analyzed, a result not found in the main reviews on the neutrino phenomenology. Designed as an introductory guide for students new to the field of Neutrino Physics, this text is supported by numerical programs available in the \href{https://github.com/GEFAN-Unicamp/Public/tree/main/ConversionProbabilities}{ GitHub repository} of our research group.

\bibliography{analytical.v2Notes}

\end{document}